\documentclass[journal,comsoc]{IEEEtran}

\ifCLASSINFOpdf
\else
\fi
\usepackage[cmex10]{amsmath}
\usepackage{amssymb}
\usepackage{siunitx}
\usepackage{caption}
\usepackage{algpseudocode}
\usepackage{algorithm}
\algnewcommand{\Initialization}[1]{%
  \State \textbf{initialization:}
  \Statex \hspace*{\algorithmicindent}\parbox[t]{.8\linewidth}{\raggedright #1}
}

\algnewcommand{\Rep}[1]{%
  \State \textbf{repeat:}
  \Statex \hspace*{\algorithmicindent}\parbox[t]{.8\linewidth}{\raggedright #1}
}

\usepackage{tikz}
\usetikzlibrary{shapes,arrows,fit,positioning,calc}
\usetikzlibrary{plotmarks}
\usetikzlibrary{decorations.pathreplacing}
\usetikzlibrary{patterns}
\usetikzlibrary{arrows,automata}

\usepackage{pgfplots}
\pgfplotsset{compat=newest}

\newcommand{\PR}[1]{\ensuremath{\left[#1\right]}}
\newcommand{\PC}[1]{\ensuremath{\left(#1\right)}}
\newcommand{\chav}[1]{\ensuremath{\left\{#1\right\}}}

\begin{document}
\title{Optimal Precoding for Multiuser MIMO Systems With Phase Quantization and PSK Modulation via Branch-and-Bound}

\author{Erico~S.~P.~Lopes,~\IEEEmembership{} and~ Lukas~T.~N.~Landau,~\IEEEmembership{Member,~IEEE,}\vspace{-1em}				
% <-this % stops a space
%\thanks{The authors would like to thank Johannes Israel from the Institute of Numerical Mathematics, TU Dresden for the introduction of the branch-and-bound method and for verifying the notation.}
\thanks{The authors are with Centro de Estudos em Telecomunica\c{c}\~{o}es Pontif\'{i}cia Universidade Cat\'{o}lica do Rio de Janeiro, Rio de Janeiro CEP 22453-900, Brazil, (email: \{erico, lukas.landau\}@cetuc.puc-rio.br).} 
\thanks{This work has been supported by the {ELIOT ANR18-CE40-0030 and FAPESP 2018/12579-7} project.}
}
\maketitle

%\vspace{-0.5em}

\begin{abstract}
MIMO systems are considered as most promising for wireless communications. However, with an increasing number of radio front ends the corresponding energy consumption and costs become an issue, which can be relieved by the utilization of low-resolution quantizers. In this study we propose an optimal precoding algorithm constrained to constant envelope signals and phase quantization that maximizes the minimum distance to the decision threshold at the receivers using a branch-and-bound strategy. The proposed algorithm is superior to the existing methods in terms of bit error rate. Numerical results show that the proposed approach has significantly lower complexity than exhaustive search.

%Multiple-antenna systems is a key technique to serve multiple users in future wireless systems.
%For low energy consumption and hardware complexity we first consider transmit symbols with constant %magnitude and then 1-bit digital-to-analog converters. We propose precoding designs which maximize %the minimum distance to the decision threshold at the receiver.
%The precoding design with 1-bit DAC corresponds to a discrete optimization problem, which we solve %exactly with a branch-and-bound strategy. We alternatively present an approximation based on %relaxation. Our results show that the proposed branch-and-bound approach has polynomial complexity. %The proposed methods outperform existing precoding methods with 1-bit DAC in terms of uncoded bit %error rate and sum-rate. The performance loss in comparison to infinite DAC resolution is small.
\end{abstract}
%\vspace{-1em}
\begin{IEEEkeywords}
Precoding, low-resolution quantization, MIMO systems, branch-and-bound methods.
\end{IEEEkeywords}



% \fi
%\IEEEpeerreviewmaketitle

\vspace{-1em}
\section{Introduction}

The increasing growth of data transmission generates a great demand for the development of high performance communication systems. One challenge in the wireless communications area is the minimization of the energy consumption without major bit error rate performance compromise. 

With this in mind, systems with low-resolution quantizers are promising, knowing that the energy consumption of data converters scales exponentially with the resolution in amplitude \cite{Walden_1999}.
%\cite{ADC_bottleneck}

Several strategies for precoding with low-resolution quantizers exist. Linear approaches such as the Zero-forcing method (ZF) \cite{ZF-Precoding} and MMSE \cite{MMSE} have a low complexity but suffer from error floor in the bit error rate. Therefore, nonlinear precoders have been designed with different design criteria.

A conventional design criterion is the MSE which is considered in the branch-and-bound (B\&B) algorithm in \cite{Jacobsson2018}.
Another widely used design criterion in given by the maximization of the minimum distance to the decision threshold (Max-Min DDT) \cite{Landau_SCC2013,Mo_2015,Landau2017,MSM_precoder}, which is promising in combination with hard detection.
In \cite{Landau2017} an optimal precoding algorithm was presented for the Max-Min DDT and 1-bit quantization at transmitter and receiver (QPSK).
In \cite{MSM_precoder} a suboptimal algorithm is developed for the Max-Min DDT criterion and $2^{q}$-PSK symbols at each transmit antenna for QAM and PSK modulation schemes.  

In the present study, we generalize the work of \cite{Landau2017} for phase quantizers with arbitrary number of phases at the transmit antennas and PSK modulation.
This extension should be considered as non trivial because in the case of PSK, each symbol cannot be decomposed in independent real and imaginary part as done in the 1-bit case.
The proposed precoder is optimal in terms of the Max-Min DDT criterion, obtained by using a sophisticated branch-and-bound strategy.
The initial step of the proposed method implies the solution of the relaxed problem subsequently rounded to the feasible set and then a tree search based algorithm is devised.

The paper is organized as follows: Section~\ref{sec:system_model} describes the system model, whereas Section~\ref{sec:precoding_task} establishes the precoder's objectives, explains the criterion and exposes the problem formulation. In Section~\ref{sec:precoder_design} the proposed precoding algorithm is described. Section~\ref{sec:numerical_results} presents and discusses numerical results, while Section VI gives the conclusions.

Regarding the notation, note that real and imaginary part operator are also applied to vectors and matrices, e.g., $\mathrm{Re}\left\{ \boldsymbol{x} \right\} = \left[ \mathrm{Re}\left\{ \left[\boldsymbol{x}\right]_1 \right\},\ldots,    \mathrm{Re}\left\{ \left[\boldsymbol{x}\right]_M \right\}  \right]^T$.
\section{System Model}
\label{sec:system_model}

\begin{figure}
%\includegraphics[width=3in]{System_Model.PNG}
%\captionsetup{justification=centering}
\input{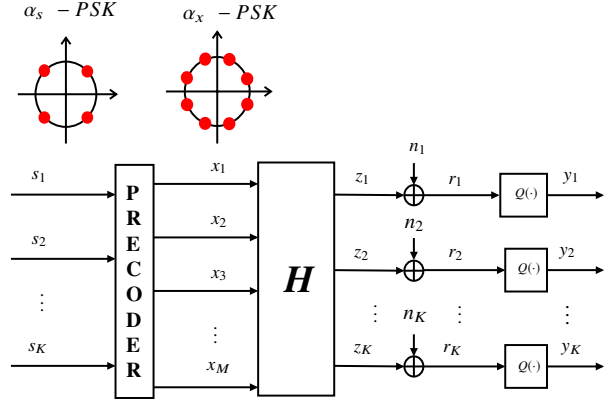}
\caption{Multiuser MIMO downlink with phase quantization and hard detection}
\label{fig:system_model}       % Give a unique label
%\vspace{-0.75em}
\end{figure}

In this study, a single cell MU-MIMO downlink with full channel state information at the base station (BS) is considered, as illustrated in Fig.~\ref{fig:system_model}. On the BS there are $M$ transmit antennas that serve $K$ single antenna users.
The data symbol for the ith user $s_i$ is a $\alpha_s$-PSK symbol taken from the set $\mathcal{S}$ described by %equation \eqref{S_set}. 
\begin{align}
	    \mathcal{S}= \left\{s: s= e^  \frac{j\pi (2 i+1) }{\alpha_{s}}  \textrm{,  for  }  i=1,\ldots, \alpha_{s} \right\}  \textrm{.}
	    \label{S_set}
	\end{align}
The stacked vector with data symbols for the $K$ users is denoted by $\boldsymbol{s}=[{s}_1,\ldots,{s}_K]^T$. %$\boldsymbol{s}_k=[{s}_{k,1},\ldots,{s}_{k,L}]^T$ for $k=1,\ldots,K$ and $l=1,\ldots,L$.
The vector $\boldsymbol{s}$ is the input for the precoder, where the transmit vector $\boldsymbol{x}=[x_1,\ldots,x_M]^T$ is constructed based on the channel.
Due to the consideration of a low-resolution data converter at the transmitter, the entries from $\boldsymbol{x}$ are constrained to the set $\mathcal{X}$, which describes an $\alpha_x$-PSK alphabet given by 
\begin{align}
	    \mathcal{X}= \left\{x: x= e^  \frac{j\pi (2 i+1) }{\alpha_{x}}  \textrm{,  for  }  i=1,\ldots, \alpha_{x} \right\} \textrm{.}
	    \label{X_set}
	\end{align}
We consider analog pulse shaping filters at the BS and matched filtering, followed by a phase quantization process at the users. Moreover, we assume perfect synchronization. In the sequel the equivalent discrete time description of the channel is considered.
A flat fading channel is considered, which is described by the matrix $\boldsymbol{H}$ whose coefficients $h_{k,m}$ are zero mean i.i.d.\ complex Gaussian random variables, where $k$ and $m$ denote the index of the user and the transmit antenna, respectively.
With this, for the noiseless case the received signals are denoted by 
\begin{align}
\label{eq:received_noiseless_sample}
z_{k}  = \sum_{m=1}^{M}     h_{k,m } \ x_m \textrm{.}
\end{align}
In the sequel a stacked vector notation is used with $\boldsymbol{z}=[{z}_1,\ldots,{z}_K]^T$.

At the receiver the signal $\boldsymbol{z}$ is corrupted by additive noise, which is denoted by the vector $\boldsymbol{n}$, which is considered to be a zero-mean i.i.d.\ complex Gaussian random vector with covariance matrix $\sigma_n^2 \boldsymbol{I}$. 

\begin{figure}
\begin{center}
    \tikzset{every picture/.style={line width=0.75pt}} %set default line width to 0.75pt        

\begin{tikzpicture}[x=0.2pt,y=0.2pt,yscale=-1,xscale=1]
%uncomment if require: \path (0,527); %set diagram left start at 0, and has height of 527

%Straight Lines [id:da5569603471012494] 
\draw    (319.5,41.25) -- (319.5,428.75) ;

\draw [shift={(319.5,39.25)}, rotate = 90] [fill={rgb, 255:red, 0; green, 0; blue, 0 }  ][line width=0.75]  [draw opacity=0] (10.72,-5.15) -- (0,0) -- (10.72,5.15) -- (7.12,0) -- cycle    ;
%Straight Lines [id:da5187048344966039] 
\draw    (512.25,234) -- (124.75,234) ;

\draw [shift={(514.25,234)}, rotate = 180] [fill={rgb, 255:red, 0; green, 0; blue, 0 }  ][line width=0.75]  [draw opacity=0] (10.72,-5.15) -- (0,0) -- (10.72,5.15) -- (7.12,0) -- cycle    ;
%Straight Lines [id:da5011038227752369] 
\draw    (450,90) -- (189,378) ;

%Straight Lines [id:da07191485870563419] 
\draw    (450,378) -- (189,90) ;

%Shape: Circle [id:dp47732513400496135] 
\draw  [color={rgb, 255:red, 247; green, 8; blue, 8 }  ,draw opacity=1 ][fill={rgb, 255:red, 250; green, 4; blue, 4 }  ,fill opacity=1 ] (446,177.5) .. controls (446,173.91) and (448.91,171) .. (452.5,171) .. controls (456.09,171) and (459,173.91) .. (459,177.5) .. controls (459,181.09) and (456.09,184) .. (452.5,184) .. controls (448.91,184) and (446,181.09) .. (446,177.5) -- cycle ;
%Shape: Circle [id:dp30967788486440884] 
\draw  [color={rgb, 255:red, 247; green, 8; blue, 8 }  ,draw opacity=1 ][fill={rgb, 255:red, 250; green, 4; blue, 4 }  ,fill opacity=1 ] (356,101.5) .. controls (356,97.91) and (358.91,95) .. (362.5,95) .. controls (366.09,95) and (369,97.91) .. (369,101.5) .. controls (369,105.09) and (366.09,108) .. (362.5,108) .. controls (358.91,108) and (356,105.09) .. (356,101.5) -- cycle ;
%Shape: Circle [id:dp0641249921142315] 
\draw  [color={rgb, 255:red, 247; green, 8; blue, 8 }  ,draw opacity=1 ][fill={rgb, 255:red, 250; green, 4; blue, 4 }  ,fill opacity=1 ] (261,101.5) .. controls (261,97.91) and (263.91,95) .. (267.5,95) .. controls (271.09,95) and (274,97.91) .. (274,101.5) .. controls (274,105.09) and (271.09,108) .. (267.5,108) .. controls (263.91,108) and (261,105.09) .. (261,101.5) -- cycle ;
%Shape: Circle [id:dp4032030012490959] 
\draw  [color={rgb, 255:red, 247; green, 8; blue, 8 }  ,draw opacity=1 ][fill={rgb, 255:red, 250; green, 4; blue, 4 }  ,fill opacity=1 ] (446,294.5) .. controls (446,290.91) and (448.91,288) .. (452.5,288) .. controls (456.09,288) and (459,290.91) .. (459,294.5) .. controls (459,298.09) and (456.09,301) .. (452.5,301) .. controls (448.91,301) and (446,298.09) .. (446,294.5) -- cycle ;
%Shape: Circle [id:dp5020010717808332] 
\draw  [color={rgb, 255:red, 247; green, 8; blue, 8 }  ,draw opacity=1 ][fill={rgb, 255:red, 250; green, 4; blue, 4 }  ,fill opacity=1 ] (365,375.5) .. controls (365,371.91) and (367.91,369) .. (371.5,369) .. controls (375.09,369) and (378,371.91) .. (378,375.5) .. controls (378,379.09) and (375.09,382) .. (371.5,382) .. controls (367.91,382) and (365,379.09) .. (365,375.5) -- cycle ;
%Shape: Circle [id:dp7448639263637076] 
\draw  [color={rgb, 255:red, 247; green, 8; blue, 8 }  ,draw opacity=1 ][fill={rgb, 255:red, 250; green, 4; blue, 4 }  ,fill opacity=1 ] (262,368.5) .. controls (262,364.91) and (264.91,362) .. (268.5,362) .. controls (272.09,362) and (275,364.91) .. (275,368.5) .. controls (275,372.09) and (272.09,375) .. (268.5,375) .. controls (264.91,375) and (262,372.09) .. (262,368.5) -- cycle ;
%Shape: Circle [id:dp4459396448726496] 
\draw  [color={rgb, 255:red, 247; green, 8; blue, 8 }  ,draw opacity=1 ][fill={rgb, 255:red, 250; green, 4; blue, 4 }  ,fill opacity=1 ] (185,285.5) .. controls (185,281.91) and (187.91,279) .. (191.5,279) .. controls (195.09,279) and (198,281.91) .. (198,285.5) .. controls (198,289.09) and (195.09,292) .. (191.5,292) .. controls (187.91,292) and (185,289.09) .. (185,285.5) -- cycle ;
%Shape: Circle [id:dp35886717034720217] 
\draw  [color={rgb, 255:red, 247; green, 8; blue, 8 }  ,draw opacity=1 ][fill={rgb, 255:red, 250; green, 4; blue, 4 }  ,fill opacity=1 ] (171,173.5) .. controls (171,169.91) and (173.91,167) .. (177.5,167) .. controls (181.09,167) and (184,169.91) .. (184,173.5) .. controls (184,177.09) and (181.09,180) .. (177.5,180) .. controls (173.91,180) and (171,177.09) .. (171,173.5) -- cycle ;
%Shape: Circle [id:dp2562257367362124] 
\draw  [color={rgb, 255:red, 4; green, 51; blue, 109 }  ,draw opacity=1 ][fill={rgb, 255:red, 4; green, 51; blue, 109 }  ,fill opacity=1 ] (483,213.5) .. controls (483,209.91) and (485.91,207) .. (489.5,207) .. controls (493.09,207) and (496,209.91) .. (496,213.5) .. controls (496,217.09) and (493.09,220) .. (489.5,220) .. controls (485.91,220) and (483,217.09) .. (483,213.5) -- cycle ;

% Text Node
\draw (370,210) node   {$\theta $};
% Text Node
\draw (465,144) node  [align=left] {$\displaystyle s_{1}$};
% Text Node
\draw (366.5,74) node  [align=left] {$\displaystyle s_{2}$};
% Text Node
\draw (246.5,78) node  [align=left] {$\displaystyle s_{3}$};
% Text Node
\draw (482.5,313) node  [align=left] {$\displaystyle s_{8}$};
% Text Node
\draw (387.5,410) node  [align=left] {$\displaystyle s_{7}$};
% Text Node
\draw (245.5,392) node  [align=left] {$\displaystyle s_{6}$};
% Text Node
\draw (155.5,276) node  [align=left] {$\displaystyle s_{5}$};
% Text Node
\draw (167.5,142) node  [align=left] {$\displaystyle s_{4}$};
% Text Node
\draw (512,201) node  [align=left] {$\displaystyle z$};

\end{tikzpicture}
    \captionsetup{justification=centering}
    \caption{Decision regions for a 8-PSK data case} 
    \label{figure:decision}
\end{center}
%\vspace{-2em}
\end{figure}
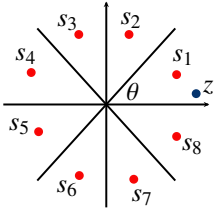

The received vector $\boldsymbol{r}$ is then given by $\boldsymbol{r} = \boldsymbol{z}+\boldsymbol{n} $, which then serves as the input for a phase quantizer which can be understood as a hard detector. %is given by the equation \eqref{eq:received_sample}.
%\boldsymbol{r} = \boldsymbol{z}+\boldsymbol{n} $,
%\begin{align}
%\label{eq:received_sample}
%\boldsymbol{r} = \boldsymbol{z}+\boldsymbol{n} 
%\end{align}
In this regard, the received signal $\boldsymbol{r}$ is elementwise uniformly phase quantized. It is considered that number of quantization regions depends on the modulation alphabet of the data ${\mathcal{S}}$ with cardinality $\alpha_s $.
%\begin{align}
%\vspace{-2em}
%\label{n_bits_phase_quantizers}
%N_b=\log_2(\alpha_s)
%\end{align}
The decision space is divided in $\alpha_s$ decision regions as is shown in Fig.~\ref{figure:decision}, one for each symbol possibility. The decision regions are circle sectors with infinite radius and angle of $2\theta$, where $\theta$ is given by $\theta=\frac{\pi}{\alpha_s}$. %The received signal $\boldsymbol{r}$ will arrive into a decision region and will be mapped to the corresponding modulation point into that region.

%\begin{align}
%\label{theta}
%\theta=\frac{\pi}{\alpha_s}
%\end{align}

The output of the phase quantizer in stacked vector notation $\boldsymbol{y}=[{y}_1,\ldots,{y}_K]^T$ is denoted by 
\begin{align}
\boldsymbol{y}=Q(\boldsymbol{r})= Q(\boldsymbol{z} + \boldsymbol{n}) =Q(  \boldsymbol{H} \boldsymbol{x} + \boldsymbol{n}) \textrm{,}
\label{detected_vector}
\end{align}
where $Q(\cdot)$ denotes the quantization operator.
Each possible output represents an element of the transmit symbol alphabet ($\boldsymbol{y} \in S^{K}$). With this, the vector $\boldsymbol{y}$ also represents the detected symbols $\hat{\boldsymbol{s}}$.   
%The received signals are applied to hard decision detectors or equivalently 1-bit ADCs described by %${y}_{k,l}=Q(z_{k,l})= \mathrm{sgn}(\mathrm{Re}\left\{z_{k,l}\right\}) +j \ %\mathrm{sgn}(\mathrm{Im}\left\{z_{k,l}\right\})$, such that ${y}_{k,l} \in \left\{ 1+j,1-j,-1+j,-1-j %\right\}$, where $Q(\cdot)$ denotes the 1-bit quantization. 
%By using vector notation the received $K L$ samples can be expressed as
%
%\begin{align}
%\boldsymbol{y}=Q(\boldsymbol{z})= Q(\boldsymbol{r} + \boldsymbol{n}) =Q(  \boldsymbol{H} %\boldsymbol{x} + \boldsymbol{n}) \textrm{,}
%\end{align}
%\noindent where $\boldsymbol{H}=[\boldsymbol{h}_1^T,\ldots,\boldsymbol{h}_K^T]$ is the channel matrix with dimensions $K \times M$ which consists of the vectors $\boldsymbol{h}_i$ each having dimensions $1 \times M$. 
\section{Precoding task}

\label{sec:precoding_task}

This section establishes the objectives of the precoder, presents the used design criterion and exposes the problem formulation.
%The main objective is to minimize the MU interference as well as the distortions due to the low-resolution quantizers. 
%In order to do that we set $\alpha_s \leq \alpha_x$ to have the vector %$\boldsymbol{x}$ with at least the same modulation order as the vector %$\boldsymbol{s}$.
The criterion for the precoder design is the maximization of the minimum distance to the decision threshold or equivalently the maximization of the safety margin at the detectors.
With this the aim is to find the vector  $\boldsymbol{x}$ which yields the $\boldsymbol{z}$ where the smallest distance to the decision threshold is maximized. By expressing the corresponding problem in the epigraph form \cite{Boyd_2004}, the problem has a linear objective function, linear constraints and a discrete feasible set, which then is a non-convex problem that has a NP hard solution by applying exhaustive search. 

%However, the proposed precoder solves it using a Branch-and-Bound algorithm which has polynomial complexity.

This study, relies on the distance to the decision threshold $\epsilon$ for hard detection of PSK symbols and the description of the objective is equivalent to the one presented in \cite{MSM_precoder}. Note that for the special case of QPSK modulation the objective is also equivalent to the objective utilized in \cite{Landau2017}.

\begin{figure}[t]
\begin{center}
    \tikzset{every picture/.style={line width=0.75pt}} %set default line width to 0.75pt        

\begin{tikzpicture}[x=0.3pt,y=0.3pt,yscale=-1,xscale=1]
%uncomment if require: \path (0,527); %set diagram left start at 0, and has height of 527

%Straight Lines [id:da6174013396145508] 
\draw    (77,64.5) -- (77,324) ;

\draw [shift={(77,62.5)}, rotate = 90] [fill={rgb, 255:red, 0; green, 0; blue, 0 }  ][line width=0.75]  [draw opacity=0] (10.72,-5.15) -- (0,0) -- (10.72,5.15) -- (7.12,0) -- cycle    ;
%Straight Lines [id:da6604459143326435] 
\draw    (281.5,260.75) -- (9,260.75) ;

\draw [shift={(283.5,260.75)}, rotate = 180] [fill={rgb, 255:red, 0; green, 0; blue, 0 }  ][line width=0.75]  [draw opacity=0] (10.72,-5.15) -- (0,0) -- (10.72,5.15) -- (7.12,0) -- cycle    ;
%Straight Lines [id:da6333718994724997] 
\draw    (152,86.5) -- (76.75,260.75) ;

%Shape: Circle [id:dp11322029634605513] 
\draw  [color={rgb, 255:red, 247; green, 8; blue, 8 }  ,draw opacity=1 ][fill={rgb, 255:red, 250; green, 4; blue, 4 }  ,fill opacity=1 ] (171,155.5) .. controls (171,151.91) and (173.91,149) .. (177.5,149) .. controls (181.09,149) and (184,151.91) .. (184,155.5) .. controls (184,159.09) and (181.09,162) .. (177.5,162) .. controls (173.91,162) and (171,159.09) .. (171,155.5) -- cycle ;
%Shape: Circle [id:dp009133621238364809] 
\draw  [color={rgb, 255:red, 4; green, 51; blue, 109 }  ,draw opacity=1 ][fill={rgb, 255:red, 4; green, 51; blue, 109 }  ,fill opacity=1 ] (152,206) .. controls (152,202.41) and (154.91,199.5) .. (158.5,199.5) .. controls (162.09,199.5) and (165,202.41) .. (165,206) .. controls (165,209.59) and (162.09,212.5) .. (158.5,212.5) .. controls (154.91,212.5) and (152,209.59) .. (152,206) -- cycle ;
%Straight Lines [id:da251070936920625] 
\draw    (251,194.5) -- (76.75,260.75) ;

%Straight Lines [id:da3051262943912103] 
\draw    (387,65) -- (387,324.5) ;

\draw [shift={(387,63)}, rotate = 90] [fill={rgb, 255:red, 0; green, 0; blue, 0 }  ][line width=0.75]  [draw opacity=0] (10.72,-5.15) -- (0,0) -- (10.72,5.15) -- (7.12,0) -- cycle    ;
%Straight Lines [id:da2169298529334227] 
\draw    (592,260.25) -- (319.5,260.25) ;

\draw [shift={(594,260.25)}, rotate = 180] [fill={rgb, 255:red, 0; green, 0; blue, 0 }  ][line width=0.75]  [draw opacity=0] (10.72,-5.15) -- (0,0) -- (10.72,5.15) -- (7.12,0) -- cycle    ;
%Straight Lines [id:da7583896795724239] 
\draw    (558,171) -- (386.92,259.89) ;

%Shape: Circle [id:dp5728666070108206] 
\draw  [color={rgb, 255:red, 247; green, 8; blue, 8 }  ,draw opacity=1 ][fill={rgb, 255:red, 250; green, 4; blue, 4 }  ,fill opacity=1 ] (531,260.5) .. controls (531,256.91) and (533.91,254) .. (537.5,254) .. controls (541.09,254) and (544,256.91) .. (544,260.5) .. controls (544,264.09) and (541.09,267) .. (537.5,267) .. controls (533.91,267) and (531,264.09) .. (531,260.5) -- cycle ;
%Shape: Circle [id:dp6575989689936119] 
\draw  [color={rgb, 255:red, 4; green, 51; blue, 109 }  ,draw opacity=1 ][fill={rgb, 255:red, 4; green, 51; blue, 109 }  ,fill opacity=1 ] (495,290.5) .. controls (495,286.91) and (497.91,284) .. (501.5,284) .. controls (505.09,284) and (508,286.91) .. (508,290.5) .. controls (508,294.09) and (505.09,297) .. (501.5,297) .. controls (497.91,297) and (495,294.09) .. (495,290.5) -- cycle ;
%Straight Lines [id:da942498063538707] 
\draw    (558,351) -- (386.92,259.89) ;

% Text Node
\draw (114.5,226) node   {$2\theta $};
% Text Node
\draw (178,132) node  [align=left] {$\displaystyle s_{i}$};
% Text Node
\draw (179.5,191) node  [align=left] {$\displaystyle z_{i}$};
% Text Node
\draw (51,71.5) node  [align=left] {$ $};
% Text Node
\draw (47,61.5) node   {$Im$};
% Text Node
\draw (268,280.5) node   {$Re$};
% Text Node
\draw (450,244) node   {$\theta $};
% Text Node
\draw (542.5,239.5) node  [align=left] {$\displaystyle s'_{i}$};
% Text Node
\draw (473,282) node  [align=left] {$\displaystyle w_{i}$};
% Text Node
\draw (361.5,71) node  [align=left] {$ $};
% Text Node
\draw (353.5,64) node   {$Im$};
% Text Node
\draw (601,277.5) node   {$Re$};

\end{tikzpicture}
    \captionsetup{justification=centering}
    \caption{ Rotated coordinate system} 
    \label{figure:Rotated coordinate system}
\end{center}
%\vspace{-1.5em}
\end{figure}
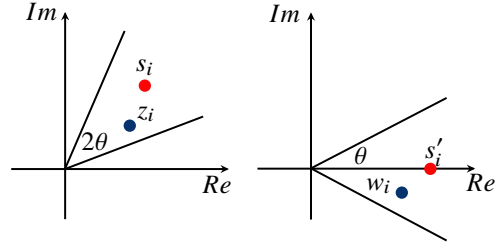

By considering a rotation by $\mathrm{arg}\{s_i^*\}=-\phi_{s_i}$ of the coordinate system  the symbol of interest is placed on the real axis, as shown in Fig.\ref{figure:Rotated coordinate system}. This is done by multiplying both the interest symbol $s_i$ and the noiseless received signal $z_i$ by $e^{-j\phi_{s_i}} = s_i^*$ which reads
\begin{equation}
\begin{aligned}
    \quad s_i^{'}=s_i s_i^*=1  \text{,}  \quad  w_i=z_i s_i^* \textrm{.}
    \label{wi and si'}
\end{aligned}
\end{equation}
The distance of the rotated symbol $w_i$ to the rotated decision threshold is then expressed as
\begin{align}
    \epsilon_i=\mathrm{Re} \left\{w_i\right\}\sin{\theta}-\lvert \mathrm{Im} \left\{w_i\right\}\rvert\cos{\theta}  \textrm{,}
    \label{epsiloni_definition}
\end{align}
as shown in detail in \cite{MSM_precoder}.
Since the considered rotation included also the decision thresholds the distance expression in \eqref{epsiloni_definition} holds also for $z_i$.
The minimum of all $\epsilon_i$, for $i=1,\ldots,M$ is defined as $\epsilon$, which serves as the objective of the precoding design. The algorithms task is to construct the transmit vector $\boldsymbol{x}$ that  maximizes $\epsilon$.

Based on a stacked vector notation for $w_i$, namely $\boldsymbol{w}  = \mathrm{diag({\boldsymbol{s^*}})}  \boldsymbol{H} \boldsymbol{x} $,
the equivalent minimization problem reads
\begin{align}
\label{eq:x_opt_first_eq}
	\begin{bmatrix}\boldsymbol{x}_{\textrm{opt}} \textrm{,}\ \epsilon_\textrm{\hspace{0.2em}opt} 
	\end{bmatrix}&=\arg\hspace{-0.75em}\min_{\boldsymbol{x} \in \mathcal{X}^M, \epsilon}  -\epsilon \ \ \ \ \ \ \textrm{s.t.}  \\
& \hspace{-2em}\mathrm{Re}\left\{\boldsymbol{H}_{s^*}\boldsymbol{x}\right\}\sin{\theta}- \vert\mathrm{Im}\left\{\boldsymbol{H}_{s^*}\boldsymbol{x}\right\}\vert \cos{\theta}\geq \epsilon\boldsymbol{1}_{2K} \notag \textrm{,}
\end{align} 
where $ \boldsymbol{H}_{s^*} = \mathrm{diag({\boldsymbol{s^*}})}  \boldsymbol{H}$.

\section{Proposed Branch-and-Bound Precoder}
\label{sec:precoder_design}

In this section we introduce the proposed precoder and derive the bounding steps for the algorithm. It is divided into three parts, the description of the mapped version of the Minimum Distance to Decision Threshold Precoder (MDDT-Mapped), a general introduction of branch-and-bound precoding strategy and the description of the MDDT branch-and-bound algorithm. 
%The main goal is to construct an algorithm for solving equation \eqref{eq:x_opt_first_eq} and finding $\boldsymbol{x}$ in a polynomial time.
\subsection{MDDT-Mapped Precoder}

One approach for finding a feasible solution of \eqref{eq:x_opt_first_eq} is to solve a relaxed version of the original problem followed by a mapping process to ensure that the precoding vector is in the feasible set of the discrete problem.

The relaxation is brought by replacing the set $\mathcal{X}^M$ by its convex hull, which then establishes convexity of the considered problem. 
The corresponding relaxed problem is an LP and reads
\begin{align}
\label{eq:lower_bound}
&	\begin{bmatrix} \boldsymbol{x}_\textrm{lb}\textrm{,}\ \epsilon_{\textrm{ lb}}
	\end{bmatrix}=  \arg\min_{\boldsymbol{x}, \epsilon}  -\epsilon  \ \ \ \ \  \textrm{s.t. }    \\
& \mathrm{Re}\left\{\boldsymbol{H}_{s^*}\boldsymbol{x}\right\}\sin{\theta}-|\mathrm{Im} \chav{\boldsymbol{H}_{s^*}\boldsymbol{x}}|\cos{\theta}\geq \epsilon\boldsymbol{1}_{2K} \notag\\
&  \mathrm{Re} \chav{{{x}}_m e^{j\phi_i}}  \leq \frac{\cos\PC{\frac{\pi}{\alpha_x}}}{\sqrt{{{M}}}} ,  \notag \textrm{  for } m=1,\ldots,M  \textrm{   and}\\
& \phi_i=\frac{2\pi i }{\alpha_x}  \textrm{,  for  }  i=1,\ldots \alpha_x  \textrm{,}\notag
\end{align} 
which is basically presented before in \cite{MSM_precoder}.
Note that unlike the algorithm in \cite{MSM_precoder}, where $\alpha_x$ is restricted to integer powers of 2, the problem formulation \eqref{eq:lower_bound} from above supports $\alpha_x$ to be any integer value.
 Subsequently the continuous solution $\boldsymbol{x}_\textrm{lb}$ is quantized to the point in $\mathcal{X}^M$ with the shortest Euclidean distance.
 %We can see in Fig.\ref{figure:Convex Hull} an example of the relaxed set for the 16-PSK case.
 
 %By relaxing the original problem and applying the discussed changes a new the relaxed version is stated as follows %on equation \eqref{eq:lower_bound}
The optimal value of \eqref{eq:lower_bound} is always a lower bound to the optimal value of the original problem \eqref{eq:x_opt_first_eq}.  Mapping to the feasible set yields a valid solution $\boldsymbol{x}_{\mathrm{ub}}$ and the corresponding value for $-\epsilon$ provides an upper bound on the optimal value of the original problem \eqref{eq:x_opt_first_eq}. 

%This corresponding design is proposed on \cite{MSM_precoder} is a sub optimal solution.
 %which will be used as the initialization of Branch and Bound. 
 %The resulting $-\epsilon$ is an upper bound of the optimal value of \eqref{eq:x_opt_first_eq}.

 %A valid solution in terms of $\boldsymbol{x} \in \mathcal{X}^M$ can be obtained by mapping the solution of \eqref{eq:lower_bound} to the discrete input vector with the smallest Euclidean distance. 

%\begin{figure}
%\begin{center}
%    \input{figures/constaints.tex}
%    \captionsetup{justification=centering}
%    \caption{Convex Hull of $\mathcal{X}$} 
%    \label{figure:Convex Hull}
%\end{center}
%\vspace{-2.5em}
%\end{figure}

%\input{Lower_bounding}
\subsection{Introduction of the Branch-and-Bound method}

This part of the algorithm is a tree search problem, where a breadth first search is employed. For constructing the tree we consider that from each node $\alpha_x$ branches goes out and that the tree consists of $M$ levels.

For the construction of the discrete precoding vector we consider a constrained minimization of a precoding objective function $f(\boldsymbol{x}, \boldsymbol{s})$, which could be the negative minimum distance to decision threshold, given by
\begin{align}
\label{eq:original_problem}
\boldsymbol{x}_{\textrm{opt}} =& \arg\min_{\boldsymbol{x}} f(\boldsymbol{x}, \boldsymbol{s} ) \ \ \textrm{    s.t. } \boldsymbol{x} \in   \mathcal{X}^{M} \textrm{.}
\end{align}
A lower bound on $f(\boldsymbol{x}_{\textrm{opt}},\boldsymbol{s})$ can be obtained by relaxing this problem, e.g., as described in \eqref{eq:lower_bound}. An upper bound on $f(\boldsymbol{x}_{\textrm{opt}},\boldsymbol{s})$ can be found by mapping the solution of the relaxed version to $\mathcal{X}^{M}$ and evaluating $f(\cdot)$ accordingly.
The upper bound on the optimal value is termed $\check{f}$. 

Note that $\check{ f } \geq  f(\boldsymbol{x}_{\textrm{opt}})$, the mapped solution, cannot yield a better solution than the relaxed solution. 
%The solution of \eqref{eq:lower_bound} is termed $\boldsymbol{x}_{\textrm{lb}}$.

If we consider $d$ fixed entries of $\boldsymbol{x}$, the precoding vector becomes $\boldsymbol{x}=[\boldsymbol{x}_1^T, \boldsymbol{x}_2^T ]^T$, with $\boldsymbol{x}_1 \in \mathcal{X}^d $.
Then a sub problem can be formulated with
\begin{align}
\label{eq:lb_subproblem}
\boldsymbol{x}_{2,\textrm{lb}} =& \arg\min_{\boldsymbol{x}_2} f(\boldsymbol{x}_2, \boldsymbol{x}_1, \boldsymbol{s} ) \\
&\textrm{s.t. }   \mathrm{Re} \chav{{x}_m e^{j\phi_i}}  \leq \frac{\cos\PC{\frac{\pi}{\alpha_x}}}{\sqrt{{{M}}}} ,  \notag \textrm{for } m=1,\ldots,M-d  \\
& \phi_i=\frac{2\pi i }{\alpha_x}  \textrm{,  for  }  i=1,\ldots \alpha_x \notag \ \textrm{.}
 \end{align}
If the optimal value of \eqref{eq:lb_subproblem} is larger (worse) than a known upper bound $\check{ f }$ on the solution of \eqref{eq:original_problem}, then all member in the discrete solution set which include vector $\boldsymbol{x}_{1}$ can be excluded from the search.
%By doing this we can exclude several candidates from the possible solution set.

%The branch-and-bound algorithm converges faster when we can compute as %early as possible an upper bound that permits many exclusions.

\vspace{-1em}
\subsection{MDDT Branch-and-Bound algorithm derivation}
\label{sec:bb_algorithm_design}

In this section a branch-and-bound algorithm is proposed which solves \eqref{eq:original_problem} by considering the problem in \eqref{eq:lower_bound} for the initialization and sub problems as given by \eqref{eq:lb_subproblem} for computing lower bounds.

 In order to formulate a real valued problem
 matrix $\boldsymbol{H}_r$ and vector $\boldsymbol{x}_r$ are defined as follows
\begin{align}
\label{eq:stacked_vector_notation1}
\quad
\boldsymbol{x}_{\textrm{r}}					=			\begin{bmatrix} \mathrm{Re} \left\{\boldsymbol{x}_1\right\} \\  \mathrm{Im} \left\{\boldsymbol{x}_1\right\} \\
\mathrm{Re} \left\{\boldsymbol{x}_2\right\} \\  \mathrm{Im} \left\{\boldsymbol{x}_2\right\} \\
\vdots \\
\mathrm{Re} \left\{\boldsymbol{x}_M\right\} \\  \mathrm{Im} \left\{\boldsymbol{x}_M\right\}
\end{bmatrix} \textrm{,} 
\quad
\boldsymbol{H}_{\textrm{r}}    &=      \mbox{\scriptsize $\begin{bmatrix} 
\Gamma_{11}  \cdots\Gamma_{1M} \\
\Lambda_{11} \cdots \Lambda_{1M}\\
\vdots\\
\Gamma_{K1}  \cdots \Gamma_{KM} \\
\Lambda_{K1}  \cdots \Lambda_{KM}\\
\Psi_{11}  \cdots\ \Psi_{1M} \\
\Delta_{11} \cdots \Delta_{1M}\\
\vdots\\
\Psi_{K1}  \cdots \Psi_{KM} \\
\Delta_{K1}  \cdots \Delta_{KM}
								\end{bmatrix}$} \text{,}
\end{align}
with
\begin{equation}
    \label{entries_of_Hr}
    \begin{aligned}
   \boldsymbol{\mathrm{\Gamma}}  & = & \mathrm{Im} \left\{  \boldsymbol{H}_{s^*}  \right\}\cos(\theta) & -\mathrm{Re} \left\{  \boldsymbol{H}_{s^*}  \right\} \sin(\theta) 
\\
    \boldsymbol{\mathrm{{\Lambda}}} & = &\mathrm{Re} \left\{  \boldsymbol{H}_{s^*}  \right\}\cos(\theta) & +\mathrm{Im} \left\{   \boldsymbol{H}_{s^*}  \right\} \sin(\theta)  
\\
          \boldsymbol{\mathrm{{\Psi}}}   & =&-  \mathrm{Im} \left\{  \boldsymbol{H}_{s^*}  \right\}\cos(\theta) & - \mathrm{Re} \left\{  \boldsymbol{H}_{s^*} \right\} \sin(\theta)  
\\
    \boldsymbol{\mathrm{{\Delta}}} & = &  \mathrm{Im} \left\{  \boldsymbol{H}_{s^*}  \right\}\sin(\theta) & -\mathrm{Re} \left\{   \boldsymbol{H}_{s^*}  \right\} \cos(\theta) \text{.}
\end{aligned}
\end{equation}
With the real valued description, the variable vector of the optimization problem can be denoted by $\boldsymbol{v}=[\epsilon,\boldsymbol{x}_{\textrm{r}}^T ]^T$ , such that the discrete optimization problem reads as 
\begin{align}
\boldsymbol{v}_{\text{opt}}=& \arg\min_{\boldsymbol{v}}  \boldsymbol{a}^T \boldsymbol{v} \\
& \textrm{s.t. } \boldsymbol{A} \boldsymbol{v} \leq  \boldsymbol{0}_{2K},  \notag \\
&     \chav{\boldsymbol{v}_{2m}+j\boldsymbol{v}_{2m+1}  }   \in \mathcal{X}_{\mathrm{}},  \ \  \textrm{for } m=1,\ldots,M     \textrm{,} \notag
\end{align}
with 
\begin{align}
\quad \boldsymbol{a}=[-1,\boldsymbol{0}_{2M}^T]^T \text{,}
\quad\boldsymbol{A}= 
\begin{bmatrix} \boldsymbol{1}_{2K} , \boldsymbol{H}_{\textrm{r}}     
\end{bmatrix}\notag\textrm{.}
\end{align}
Replacing the the discrete solution set by its convex hull yields the relaxed problem given by
\begin{align}
\label{eq:v_real}
\boldsymbol{v}_{\text{lb}}=& \arg\min_{\boldsymbol{v}}  \boldsymbol{a}^T \boldsymbol{v} \ \ \ \textrm{s.t. } \boldsymbol{U} \boldsymbol{v} \leq  \boldsymbol{p}_{},  
\end{align}
with
\begin{align}
\quad &\boldsymbol{U}= \begin{bmatrix} \boldsymbol{A}^T,\boldsymbol{R}^T \end{bmatrix}^T 
\quad\boldsymbol{R}=  \begin{bmatrix} \boldsymbol{0}_{M\alpha_x} , \boldsymbol{R}' \end{bmatrix} \notag
\end{align}
%\vspace{-2em}
\begin{align}
\notag &\boldsymbol{R}'=\begin{bmatrix} (\boldsymbol{I}_M\otimes \boldsymbol{\beta}_1)^T, (\boldsymbol{I}_M\otimes \boldsymbol{\beta}_2)^T, \ldots ,(\boldsymbol{I}_M\otimes \boldsymbol{\beta}_{\alpha_x})^T \end{bmatrix}^T \notag
\end{align}
\vspace{-2em}
\begin{align}
&\quad \boldsymbol{\beta}_i=\begin{bmatrix} \cos{\phi_i}, - \sin{\phi_i}\end{bmatrix} 
\quad \boldsymbol{p}=  \begin{bmatrix} \boldsymbol{0}_{2K} , \frac{\cos(\frac{\pi}{\alpha_x})}{\sqrt{{{M}}}} \boldsymbol{1}_{M\alpha_x}  \end{bmatrix}^T \notag \textrm{.}
\end{align} 
In the branch-and-bound method sub problems are solved due to $\boldsymbol{v}=\left[\epsilon,  \boldsymbol{x}_{r_1}^T,\boldsymbol{x}_{r_2}^T   \right]^T$, where $\boldsymbol{x}_{r_1}$ is a fixed vector of length $2d$, which belongs to the discrete set according to ${\boldsymbol{v}_{1_{2m}}+j\boldsymbol{v}_{1_{2m+1}}  }   \in \mathcal{X}_{\mathrm{}},  \ \  \textrm{for } m=1,\ldots,d$.

The matrix $\boldsymbol{U}$ can be expressed with the following structure $\boldsymbol{U}=\PR{\boldsymbol{u}_1, \boldsymbol{U}_1,  \boldsymbol{U}_2   }$, where $\boldsymbol{U}_1$ contains $2d$ columns of $\boldsymbol{U}$ and $\boldsymbol{u}_1$ is the first column of $\boldsymbol{U}$. With this, the matrix $\Tilde{\boldsymbol{U}}= \begin{bmatrix}\boldsymbol{u}_1 ,\boldsymbol{U}_2 \end{bmatrix}$ and the vector $\Tilde{\boldsymbol{v}}= \begin{bmatrix}\epsilon,  \boldsymbol{x}_{r_2}^T \end{bmatrix}^T$ are composed. Using $\Tilde{\boldsymbol{U}}$ and $\Tilde{\boldsymbol{v}}$ the sub problem for the lower-bounding step can be expressed as
\begin{align}
\label{eq:bb_optimzation_problem}
{\Tilde{\boldsymbol{v}}_{\textrm{lb}}}=& \arg\min_{\Tilde{\boldsymbol{v}}}  \Tilde{\boldsymbol{a}}^T \Tilde{\boldsymbol{v}} \ \ \ \textrm{s.t. } \Tilde{\boldsymbol{U}} \Tilde{\boldsymbol{v}} \leq  \boldsymbol{b} \textrm{,}
\end{align}   
with $\Tilde{\boldsymbol{a}}=\left[-1, \boldsymbol{0}_{2M-2d}^T \right]^T$ and $\boldsymbol{b}=\boldsymbol{p}-\boldsymbol{U}_1 \boldsymbol{x}_{r_1}$. 
Solving \eqref{eq:bb_optimzation_problem} provides an upper bound on the optimal value of the discrete problem with the condition on $\boldsymbol{x}_{r_1}$.
In case the lower bound conditioned on $\boldsymbol{x}_{r_1}$ is higher than any upper bound on the original problem $\boldsymbol{x}_{r_1}$ cannot be part of the solution and every member of the discrete solution set which includes $\boldsymbol{x}_{r_1}$ can be excluded from the search.
The steps of the method are detailed in Algorithm~\ref{alg:BB_Precoding}. 

%The sub problem \eqref{eq:bb_optimzation_problem} is an LP that can be %solved by active set methods, which can take advantage of initialization %vectors near the optimum. This can be practically exploited by the %branch-and-bound strategy where a series of similar problems have to be %solved.
\begin{algorithm}
  \caption{Proposed B\&B Precoding for solving \eqref{eq:x_opt_first_eq}}
	\label{alg:BB_Precoding}
  \begin{algorithmic}    %[1] %for numbers
	\Initialization{}
		\vspace{-1.25em}
	\State{Given the channel $\boldsymbol{H}$ and transmit symbols $\boldsymbol{s}$ compute a valid upper bound $\check{f}$ on the problem in \eqref{eq:x_opt_first_eq}, e.g., by solving \eqref{eq:lower_bound} followed by a mapping to the closest precoding vector $\boldsymbol{x} \in \mathcal{X}_{\textrm{}}^{M}$}
	\vspace{2mm}
	\State{Define the first level ($d=1$) of the tree by $\mathcal{G}_{d}:=\mathcal{X}_{\textrm{}}$}
	\vspace{2mm}	
	\For{$d=1:M-1$}
	\State{ Partition  $\mathcal{G}_{d}$ in $\boldsymbol{x}_{1,1},\ldots,\boldsymbol{x}_{1,\left|\mathcal{G}_{d}\right|}$ }  
	
	  \For{$i=1:\left| \mathcal{G}_{d} \right|$}
		\vspace{2mm}
		\State{Express $\boldsymbol{x}_{1,i}$ with stacked vector notation  }
		\State{due to \eqref{eq:stacked_vector_notation1}  as $\boldsymbol{x_r}_{1,i}$  }
		\State{Conditioned on $\boldsymbol{x_r}_{1,i}$ solve $\Tilde{\boldsymbol{v}}_{\textrm{lb}}$ from \eqref{eq:bb_optimzation_problem}  }
		\State{Determine $\epsilon=\left[\Tilde{\boldsymbol{v}}_{\textrm{lb}}\right]_{1}$}
		\State{Compute the lower bound:  $\mathrm{lb}(\boldsymbol{x}_{1,i}):=  -\epsilon $;}
		\vspace{2mm}
		\State{Map $\boldsymbol{x}_{2,\mathrm{lb}}$ to the discrete solution with the closest} 
	  \State{Euclidean distance:}
		\State{$\check{\boldsymbol{x}}_2(\boldsymbol{x}_{2,\mathrm{lb}}) \in \mathcal{X}_{\textrm{}}^{M-d} $}
		\State{Using $\check{\boldsymbol{x}}_2$ find the smallest (negative) distance to the}
		\State{decision threshold  $\mathrm{ub}(\boldsymbol{x}_{1,i}) :=$ }	
		\begin{align*}
		 \max_{k}  \left[ 
		\bigg| \mathrm{Im} \biggl\{ \boldsymbol{H}_{s^*}  
					        \begin{bmatrix} {\boldsymbol{x}_{1,i}} \\   {\check{\boldsymbol{x}}_{2}}   \end{bmatrix}
					        \biggr\}\bigg|
					        \cos{\theta} \rbrace  
		-  \mathrm{Re} \biggl\{ \boldsymbol{H}_{s^*}  
					        \begin{bmatrix} {\boldsymbol{x}_{1,i}} \\   {\check{\boldsymbol{x}}_{2}}   \end{bmatrix}\biggr\} \sin{\theta}       \right]_k
									\end{align*}  
									\State{Update the best upper bound with:}
		\State{$\check{f} =\min\left( \check{f}, \mathrm{ub}(\boldsymbol{x}_{1,i})  \right)    $}
	\EndFor
	\State{Build a reduced set by comparing conditioned  }
	\State{lower bounds with the global upper bound $\check{f}$}
	\State{$\mathcal{G}_{d}^{\prime}:=\left\{  \boldsymbol{x}_{2,i} \vert \mathrm{lb}(\boldsymbol{x}_{2,i})  \leq  \check{f}       , i=1,\ldots,  \left|\mathcal{G}_{d}\right|  \right\} $}
	\vspace{2mm}
	\State{Define the set for the next level in the tree}
	\State{$\mathcal{G}_{d+1}:=\mathcal{G}_{d}^{\prime} \times \mathcal{X}_{\textrm{}}$}
	\EndFor
	\State{ Search method for the ultimate level $d=M$,}
	\State{ Partition $\mathcal{G}_{1}$ in $\boldsymbol{x}_{1,1},\ldots,\boldsymbol{x}_{1,\left|\mathcal{G}_{1}\right|}$ }	
	\vspace{-1.5em}
	  \State{  \begin{align*}
		\epsilon(\boldsymbol{x}_{1,i}) := \min_{k}   \left[  
		\mathrm{Re}\left\{\boldsymbol{H}_{s^*}\boldsymbol{x}_{1,i}\right\}\sin{\theta}-|\mathrm{Im} \chav{\boldsymbol{H}_{s^*}\boldsymbol{x}_{1,i}}|\cos{\theta}
		\right]_k    \end{align*}}	
\State{The global solution is 
	\vspace{-0.8em}
\begin{align*}
 \boldsymbol{x}_{\textrm{opt}} = \mathrm{arg} \hspace{-0.60em}\max_{\boldsymbol{x}_{1,i} \in \mathcal{G}_{1}} \epsilon(\boldsymbol{x}_{1,i})  		
\end{align*}}
\end{algorithmic}
\end{algorithm}
Note that the computation of the optimal precoding vector in each symbol period can correspond to an enormous computational complexity. Nevertheless, the method might be a practical solution for channels with large coherence time, where the finite number of different precoding vectors can be precomputed and stored as suggested in \cite{Jedda_2016}.

%Computing the precoding vector in each time instance exceeds the %computational capacities in most applications. However, for small arrays %and a large coherence time of the channel, a look-up-table can store the %set of precoding vectors as suggested in \cite{Jedda_2016}. Considering %the symmetries of the constellation $\alpha_s^{K -1}$ precoding vectors %need to be computed and stored.

\section{Numerical Results}
\label{sec:numerical_results}

For comparison of the proposed method with the state-of-the-art algorithms, the uncoded bit error rate is evaluated, where Gray-coding is considered.
The considered signal-to-noise ratios (SNR) is defined by $\mathrm{SNR}=\frac{ \left\|\boldsymbol{x}\right\|^2_2 }{\sigma_n^2}$,
where $N_0$ denotes the noise power density.

%\frac{\mathrm{E}\left\{ E_{\textrm{Tx}}  \right\}   }{N_0} =

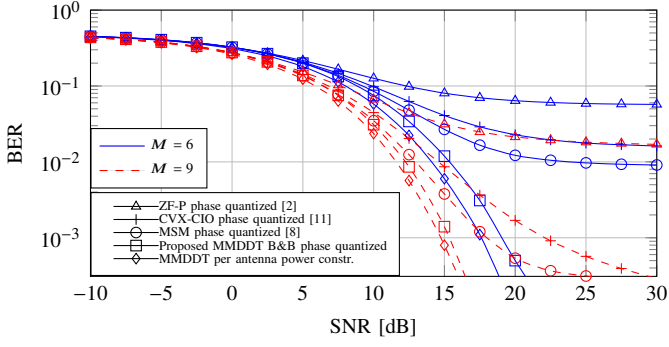
\begin{figure}[t]
%\vspace{-2em}
\begin{center}
%\sidecaption[t]
% This file was created by matlab2tikz v0.4.7 running on MATLAB 8.3.
% Copyright (c) 2008--2014, Nico Schlömer <nico.schloemer@gmail.com>
% All rights reserved.
% Minimal pgfplots version: 1.3
% 
% The latest updates can be retrieved from
%   http://www.mathworks.com/matlabcentral/fileexchange/22022-matlab2tikz
% where you can also make suggestions and rate matlab2tikz.
% 
%
% defining custom colors
\usetikzlibrary{positioning,calc}

\definecolor{mycolor1}{rgb}{0.00000,1.00000,1.00000}%
\definecolor{mycolor2}{rgb}{1.00000,0.00000,1.00000}%

\pgfplotsset{every axis label/.append style={font=\footnotesize},
every tick label/.append style={font=\footnotesize}
}

\begin{tikzpicture}[font=\footnotesize] 

\begin{axis}[%
name=ber,
ymode=log,
width  = 0.85\columnwidth,%5.63489583333333in,
%height = 0.3\columnwidth,%4.16838541666667in,
height = 0.4\columnwidth,%4.16838541666667in,
scale only axis,
xmin  = -10,
xmax  = 30,
xlabel= {SNR  [dB]},
xmajorgrids,
ymin=0.00031,
ymax=1,
ylabel={BER},
ymajorgrids,
legend entries={$M=6$,
				$M=9$,
                %$M=14$,
                %$M=16$,
                },
legend style={at={(0,0.33)},anchor=south west,draw=black,fill=white,legend cell align=left,font=\tiny}
]

\addlegendimage{solid,no marks,color=blue,fill=gray!20,mark=square}
%\addlegendimage{only marks,color=black,fill=green!20,mark=triangle}
%\addlegendimage{only marks,color=black,fill=red!20,mark=o}
\addlegendimage{dashed,no marks,color=red,fill=gray!20,mark=diamond}

%FOR THE FIRST VALUE OF M
%PoPQ
%\addplot+[smooth,color=green,solid, every mark/.append style={solid, fill=gray!20},mark=square,
%y filter/.code={\pgfmathparse{\pgfmathresult-0}\pgfmathresult}]
%  table[row sep=crcr]{%
%   0  0.013200856250000\\
%   1  0.012471784615385\\
%   2  0.011701737019231\\
%   3  0.010904155769231\\
%   4  0.010084902884615\\
%   5  0.009252383173077\\
%   6  0.008428209615385\\
%   7  0.007620877884615\\
%   8  0.006853329807692\\
%   9  0.006140354807692\\
%   10 0.005483773557692\\
%   11 0.004902665384615\\
%   12 0.004395031730769\\
%   13 0.003964064423077\\
%   14 0.003599219711538\\
%   15 0.003298030769231\\
%   16 0.003057620673077\\
%   17 0.002862022115385\\
%   18 0.002710360576923\\
%   19 0.002588457211538\\
%   20 0.002494077884615\\
%   21 0.002421959134615\\
%   22 0.002364371634615\\
%   23 0.002320342788462\\
%   24 0.002284522115385\\
%   25 0.002255893750000\\
%};

% max min bb
\addplot+[smooth,color=blue,solid, every mark/.append style={solid, fill=blue!20},mark=square,
y filter/.code={\pgfmathparse{\pgfmathresult-0}\pgfmathresult}]
  table[row sep=crcr]{%
-10	0.4517	\\
-7.5	0.4332	\\
-5	0.4074	\\
-2.5	0.3719	\\
0	0.3249	\\
2.5	0.2664	\\
5	0.2001	\\
7.5	0.1332	\\
10	0.0751	\\
12.5	0.0341	\\
15	0.0119	\\
17.5	0.0031	\\
20	0.0005	\\
22.5	0.0001	\\
%25	0	\\
%27.5	0	\\
%30	0	\\
};

%max min continuous with power constraint
\addplot+[smooth,color=blue,solid, every mark/.append style={solid, fill=cyan!20},mark=diamond,
y filter/.code={\pgfmathparse{\pgfmathresult-0}\pgfmathresult}]
  table[row sep=crcr]{%
-10	0.4477	\\
-7.5	0.4276	\\
-5	0.3996	\\
-2.5	0.3612	\\
0	0.3106	\\
2.5	0.2487	\\
5	0.1801	\\
7.5	0.113	\\
10	0.0578	\\
12.5	0.0223	\\
15	0.006	\\
17.5	0.0011	\\
20	0.0001	\\
%22.5	0	\\
%25	0	\\
%27.5	0	\\
%30	0	\\
};

%ZF
\addplot+[smooth,color=blue,solid, every mark/.append style={solid, fill=green!20},mark=triangle,
y filter/.code={\pgfmathparse{\pgfmathresult-0}\pgfmathresult}]
  table[row sep=crcr]{%
-10	0.4493	\\
-7.5	0.4301	\\
-5	0.4038	\\
-2.5	0.3684	\\
0	0.3235	\\
2.5	0.271	\\
5	0.2163	\\
7.5	0.1662	\\
10	0.1263	\\
12.5	0.0982	\\
15	0.0804	\\
17.5	0.07	\\
20	0.064	\\
22.5	0.0606	\\
25	0.0587	\\
27.5	0.0577	\\
30	0.0572	\\
};

% MSM
\addplot+[smooth,color=blue,solid, every mark/.append style={solid, fill=blue!50},mark=o,
y filter/.code={\pgfmathparse{\pgfmathresult-0}\pgfmathresult}]
  table[row sep=crcr]{%
-10	0.4506	\\
-7.5	0.4317	\\
-5	0.4055	\\
-2.5	0.3696	\\
0	0.3225	\\
2.5	0.2647	\\
5	0.2008	\\
7.5	0.1381	\\
10	0.0852	\\
12.5	0.0481	\\
15	0.0268	\\
17.5	0.0166	\\
20	0.0122	\\
22.5	0.0105	\\
25	0.0097	\\
27.5	0.0093	\\
30	0.0091	\\
};

% CVX-CIO
\addplot+[smooth,color=blue,solid, every mark/.append style={solid, fill=blue!50},mark=+,
y filter/.code={\pgfmathparse{\pgfmathresult-0}\pgfmathresult}]
  table[row sep=crcr]{%
-10	0.4501	\\
-7.5	0.4312	\\
-5	0.4049	\\
-2.5	0.3691	\\
0	0.3226	\\
2.5	0.2665	\\
5	0.2054	\\
7.5	0.1467	\\
10	0.0977	\\
12.5	0.0626	\\
15	0.0408	\\
17.5	0.029	\\
20	0.0228	\\
22.5	0.0195	\\
25	0.0177	\\
27.5	0.0168	\\
30	0.0162	\\
};

%FOR THE SECOND VALUE OF M

% max min bb
\addplot+[smooth,color=red,dashed, every mark/.append style={solid, fill=blue!20},mark=square,
y filter/.code={\pgfmathparse{\pgfmathresult-0}\pgfmathresult}]
  table[row sep=crcr]{%
-10	0.4372	\\
-7.5	0.4129	\\
-5	0.3792	\\
-2.5	0.3337	\\
0	0.2757	\\
2.5	0.208	\\
5	0.1378	\\
7.5	0.0753	\\
10	0.0309	\\
12.5	0.0086	\\
15	0.0014	\\
17.5	0.0001	\\
%20	0	\\
%22.5	0	\\
%25	0	\\
%27.5	0	\\
%30	0	\\
};

%max min continuous with power constraint
\addplot+[smooth,color=red,dashed, every mark/.append style={solid, fill=cyan!20},mark=diamond,
y filter/.code={\pgfmathparse{\pgfmathresult-0}\pgfmathresult}]
  table[row sep=crcr]{%
-10	0.43289405	\\
-7.5	0.406876283	\\
-5	0.370844067	\\
-2.5	0.322603656	\\
0	0.262044592	\\
2.5	0.192844096	\\
5	0.122988842	\\
7.5	0.063296723	\\
10	0.023625063	\\
12.5	0.00567486	\\
15	0.000797363	\\
17.5	5.70E-05	\\
20	1.31E-06	\\
22.5	2.08E-09	\\
%25	0	\\
%27.5	0	\\
%30	0	\\
};

%ZF
\addplot+[smooth,color=red,dashed, every mark/.append style={solid, fill=green!20},mark=triangle,
y filter/.code={\pgfmathparse{\pgfmathresult-0}\pgfmathresult}]
  table[row sep=crcr]{%
-10	0.4343	\\
-7.5	0.4091	\\
-5	0.3746	\\
-2.5	0.3292	\\
0	0.2736	\\
2.5	0.2123	\\
5	0.1528	\\
7.5	0.1027	\\
10	0.0666	\\
12.5	0.0441	\\
15	0.0314	\\
17.5	0.0248	\\
20	0.0213	\\
22.5	0.0194	\\
25	0.0182	\\
27.5	0.0175	\\
30	0.017	\\
};

% MSM

\addplot+[smooth,color=red,dashed, every mark/.append style={solid, fill=blue!50},mark=o,
y filter/.code={\pgfmathparse{\pgfmathresult-0}\pgfmathresult}]
  table[row sep=crcr]{%
-10	0.435598354	\\
-7.5	0.410716325	\\
-5	0.376250338	\\
-2.5	0.329989406	\\
0	0.271684542	\\
2.5	0.204695898	\\
5	0.136565006	\\
7.5	0.077072823	\\
10	0.034911056	\\
12.5	0.012451338	\\
15	0.003778398	\\
17.5	0.001201323	\\
20	0.000538148	\\
22.5	0.000367735	\\
25	0.000313504	\\
27.5	0.000286577	\\
30	0.000267167	\\
};

% CVX-CIO
\addplot+[smooth,color=red,dashed, every mark/.append style={solid, fill=blue!50},mark=+,
y filter/.code={\pgfmathparse{\pgfmathresult-0}\pgfmathresult}]
  table[row sep=crcr]{%
-10	0.435340117	\\
-7.5	0.410430715	\\
-5	0.376056092	\\
-2.5	0.330153875	\\
0	0.272793838	\\
2.5	0.20767235	\\
5	0.14233029	\\
7.5	0.085736748	\\
10	0.044626508	\\
12.5	0.020372481	\\
15	0.008627985	\\
17.5	0.003655246	\\
20	0.001695088	\\
22.5	0.00091281	\\
25	0.000568921	\\
27.5	0.000393571	\\
30	0.000290142	\\
};

%\addplot[smooth,color=green,solid,mark=none,
%y filter/.code={\pgfmathparse{\pgfmathresult-0}\pgfmathresult}]
%  table[row sep=crcr]{%
%	1 2\\
%};\label{P0}

\addplot[smooth,color=black,solid,mark=square,
y filter/.code={\pgfmathparse{\pgfmathresult-0}\pgfmathresult}]
  table[row sep=crcr]{%
	1 2\\
};\label{P1}

\addplot[smooth,color=black,solid, mark=diamond,
y filter/.code={\pgfmathparse{\pgfmathresult-0}\pgfmathresult}]
  table[row sep=crcr]{%
	1 2\\
};\label{P2}
\addplot[smooth,color=black,solid,mark=+
,
y filter/.code={\pgfmathparse{\pgfmathresult-0}\pgfmathresult}]
  table[row sep=crcr]{%
	1 2\\
};\label{P3}

\addplot[smooth,color=black,solid,mark=o,
y filter/.code={\pgfmathparse{\pgfmathresult-0}\pgfmathresult}]
  table[row sep=crcr]{%
	1 2\\
};\label{P4}
%
%\addplot[smooth,color=gray,solid,mark=none,
%y filter/.code={\pgfmathparse{\pgfmathresult-0}\pgfmathresult}]
%  table[row sep=crcr]{%
%	1 2\\
%};\label{P5}

\addplot[smooth,color=black,solid,mark=triangle,
y filter/.code={\pgfmathparse{\pgfmathresult-0}\pgfmathresult}]
  table[row sep=crcr]{%
	1 2\\
};\label{P8}

\node [draw,fill=white,font=\tiny,anchor= north  west] at (axis cs: -10,0.004) {
\setlength{\tabcolsep}{0.5mm}
\renewcommand{\arraystretch}{.8}
\begin{tabular}{l}
%\ref{P0}{Max Min total power constr.~\cite{Mo_2015}}\\
%\ref{P0}{PoPQ}\\
%\ref{P6}{Proposed PoP-8-PSK}\\
\ref{P8}{ZF-P phase quantized \cite{ZF-Precoding}}\\
\ref{P3}{CVX-CIO phase quantized \cite{CVX-CIO}}\\
\ref{P4}{MSM phase quantized \cite{MSM_precoder}}\\
\ref{P1}{Proposed MMDDT B\&B phase quantized}\\
\ref{P2}{MMDDT per antenna power constr.}\\
\end{tabular}
};

\end{axis}

\end{tikzpicture}%
%\vspace{-0.5em}
\captionsetup{justification=centering}
\caption{Uncoded BER versus $\mathrm{SNR}$, $K=2$, $\alpha_s=8$ and $\alpha_x=8$} 
\label{fig:BER_2}       % Give a unique label
%\vspace{-1em}
\end{center}
\end{figure}

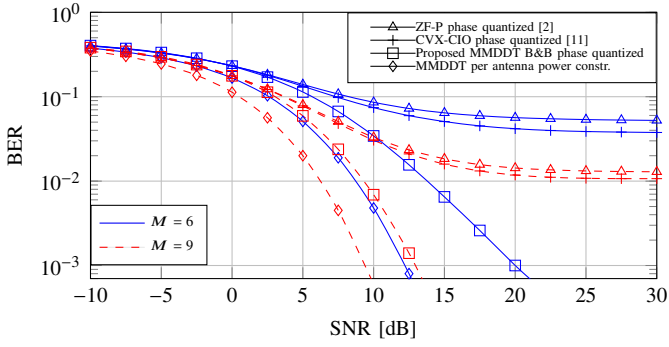
\begin{figure}[t]
%\vspace{-2em}
\begin{center}
%\sidecaption[t]
% This file was created by matlab2tikz v0.4.7 running on MATLAB 8.3.
% Copyright (c) 2008--2014, Nico Schlömer <nico.schloemer@gmail.com>
% All rights reserved.
% Minimal pgfplots version: 1.3
% 
% The latest updates can be retrieved from
%   http://www.mathworks.com/matlabcentral/fileexchange/22022-matlab2tikz
% where you can also make suggestions and rate matlab2tikz.
% 
%
% defining custom colors
\usetikzlibrary{positioning,calc}

\definecolor{mycolor1}{rgb}{0.00000,1.00000,1.00000}%
\definecolor{mycolor2}{rgb}{1.00000,0.00000,1.00000}%

\pgfplotsset{every axis label/.append style={font=\footnotesize},
every tick label/.append style={font=\footnotesize}
}

\begin{tikzpicture}[font=\footnotesize] 

\begin{axis}[%
name=ber,
ymode=log,
width  = 0.85\columnwidth,%5.63489583333333in,
%height = 0.3\columnwidth,%4.16838541666667in,
height = 0.4\columnwidth,%4.16838541666667in,
scale only axis,
xmin  = -10,
xmax  = 30,
xlabel= {SNR  [dB]},
xmajorgrids,
ymin=0.0007,
ymax=1,
ylabel={BER},
ymajorgrids,
legend entries={$M=6$,
				$M=9$,
                %$M=14$,
                %$M=16$,
                },
legend style={at={(0,0.05)},anchor=south west,draw=black,fill=white,legend cell align=left,font=\tiny}
]

\addlegendimage{solid,no marks,color=blue,fill=gray!20,mark=square}
%\addlegendimage{only marks,color=black,fill=green!20,mark=triangle}
%\addlegendimage{only marks,color=black,fill=red!20,mark=o}
\addlegendimage{dashed,no marks,color=red,fill=gray!20,mark=diamond}

%FOR THE FIRST VALUE OF M
%PoPQ
%\addplot+[smooth,color=green,solid, every mark/.append style={solid, fill=gray!20},mark=square,
%y filter/.code={\pgfmathparse{\pgfmathresult-0}\pgfmathresult}]
%  table[row sep=crcr]{%
%   0  0.013200856250000\\
%   1  0.012471784615385\\
%   2  0.011701737019231\\
%   3  0.010904155769231\\
%   4  0.010084902884615\\
%   5  0.009252383173077\\
%   6  0.008428209615385\\
%   7  0.007620877884615\\
%   8  0.006853329807692\\
%   9  0.006140354807692\\
%   10 0.005483773557692\\
%   11 0.004902665384615\\
%   12 0.004395031730769\\
%   13 0.003964064423077\\
%   14 0.003599219711538\\
%   15 0.003298030769231\\
%   16 0.003057620673077\\
%   17 0.002862022115385\\
%   18 0.002710360576923\\
%   19 0.002588457211538\\
%   20 0.002494077884615\\
%   21 0.002421959134615\\
%   22 0.002364371634615\\
%   23 0.002320342788462\\
%   24 0.002284522115385\\
%   25 0.002255893750000\\
%};

% CVX-CIO
\addplot+[smooth,color=blue,solid, every mark/.append style={solid, fill=blue!50},mark=+,
y filter/.code={\pgfmathparse{\pgfmathresult-0}\pgfmathresult}]
  table[row sep=crcr]{%
-10	0.4005	\\
-7.5	0.3693	\\
-5	0.3305	\\
-2.5	0.2839	\\
0	0.2321	\\
2.5	0.1796	\\
5	0.1332	\\
7.5	0.0978	\\
10	0.0743	\\
12.5	0.0597	\\
15	0.0508	\\
17.5	0.0451	\\
20	0.0418	\\
22.5	0.0398	\\
25	0.0387	\\
27.5	0.0381	\\
30	0.0377	\\
};

% max min bb
\addplot+[smooth,color=blue,solid, every mark/.append style={solid, fill=blue!20},mark=square,
y filter/.code={\pgfmathparse{\pgfmathresult-0}\pgfmathresult}]
  table[row sep=crcr]{%
-10	0.4032	\\
-7.5	0.3725	\\
-5	0.3335	\\
-2.5	0.2857	\\
0	0.2302	\\
2.5	0.1707	\\
5	0.1138	\\
7.5	0.0669	\\
10	0.0343	\\
12.5	0.0156	\\
15	0.0065	\\
17.5	0.0026	\\
20	0.001	\\
22.5	0.0004	\\
25	0.0001	\\
%27.5	0	\\
%30	0	\\
};

%max min continuous with power constraint
\addplot+[smooth,color=blue,solid, every mark/.append style={solid, fill=cyan!20},mark=diamond,
y filter/.code={\pgfmathparse{\pgfmathresult-0}\pgfmathresult}]
  table[row sep=crcr]{%
-10	0.3767	\\
-7.5	0.3379	\\
-5	0.2893	\\
-2.5	0.2311	\\
0	0.1664	\\
2.5	0.1027	\\
5	0.0510	\\
7.5	0.0189	\\
10	0.0048	\\
12.5	0.0008	\\
15	0.0001	\\
%17.5	0.0011	\\
%20	0.0001	\\
%22.5	0	\\
%25	0	\\
%27.5	0	\\
%30	0	\\
};

%ZF
\addplot+[smooth,color=blue,solid, every mark/.append style={solid, fill=green!20},mark=triangle,
y filter/.code={\pgfmathparse{\pgfmathresult-0}\pgfmathresult}]
  table[row sep=crcr]{%
-10	0.3989	\\
-7.5	0.3675	\\
-5	0.3287	\\
-2.5	0.2827	\\
0	0.2323	\\
2.5	0.1825	\\
5	0.1396	\\
7.5	0.1074	\\
10	0.0859	\\
12.5	0.0725	\\
15	0.0644	\\
17.5	0.0595	\\
20	0.0565	\\
22.5	0.0548	\\
25	0.0536	\\
27.5	0.0529	\\
30	0.0524	\\
};

% MSM
%\addplot+[smooth,color=blue,solid, every mark/.append style={solid, %fill=blue!50},mark=o,
%y filter/.code={\pgfmathparse{\pgfmathresult-0}\pgfmathresult}]
%  table[row sep=crcr]{%
%-10	0.4113	\\
%-7.5	0.382	\\
%-5	0.3449	\\
%-2.5	0.3002	\\
%0	0.2495	\\
%2.5	0.1971	\\
%5	0.1489	\\
%7.5	0.11	\\
%10	0.0822	\\
%12.5	0.064	\\
%15	0.0525	\\
%17.5	0.0458	\\
%20	0.0422	\\
%22.5	0.0404	\\
%25	0.0395	\\
%27.5	0.0391	\\
%30	0.0391	\\
%};

%FOR THE SECOND VALUE OF M

% CVX-CIO
\addplot+[smooth,color=red,dashed, every mark/.append style={solid, fill=blue!50},mark=+,
y filter/.code={\pgfmathparse{\pgfmathresult-0}\pgfmathresult}]
  table[row sep=crcr]{%
-10	0.375	\\
-7.5	0.3365	\\
-5	0.2895	\\
-2.5	0.2348	\\
0	0.1766	\\
2.5	0.1218	\\
5	0.0778	\\
7.5	0.048	\\
10	0.0305	\\
12.5	0.021	\\
15	0.0159	\\
17.5	0.0132	\\
20	0.0118	\\
22.5	0.0111	\\
25	0.0108	\\
27.5	0.0107	\\
30	0.0107	\\
};

% max min bb
\addplot+[smooth,color=red,dashed, every mark/.append style={solid, fill=blue!20},mark=square,
y filter/.code={\pgfmathparse{\pgfmathresult-0}\pgfmathresult}]
  table[row sep=crcr]{%
-10	0.3813	\\
-7.5	0.3439	\\
-5	0.297	\\
-2.5	0.2404	\\
0	0.1769	\\
2.5	0.1131	\\
5	0.0593	\\
7.5	0.0238	\\
10	0.0069	\\
12.5	0.0014	\\
15	0.0002	\\
%17.5	0	\\
%20	0	\\
%22.5	0	\\
%25	0	\\
%27.5	0	\\
%30	0	\\
};

%max min continuous with power constraint
\addplot+[smooth,color=red,dashed, every mark/.append style={solid, fill=cyan!20},mark=diamond,
y filter/.code={\pgfmathparse{\pgfmathresult-0}\pgfmathresult}]
  table[row sep=crcr]{%
-10	0.3478	\\
-7.5	0.3012	\\
-5	0.2444	\\
-2.5	0.1794	\\
0	0.1130	\\
2.5	0.0563	\\
5	0.0200	\\
7.5	0.0045	\\
10	0.0006	\\
%12.5	0.0000	\\
%15	0	\\
%17.5	0	\\
%20	0	\\
%22.5	0	\\
%25	0	\\
%27.5	0	\\
%30	0	\\
};

%ZF
\addplot+[smooth,color=red,dashed, every mark/.append style={solid, fill=green!20},mark=triangle,
y filter/.code={\pgfmathparse{\pgfmathresult-0}\pgfmathresult}]
  table[row sep=crcr]{%
-10	0.3737	\\
-7.5	0.3351	\\
-5	0.2881	\\
-2.5	0.2337	\\
0	0.1765	\\
2.5	0.123	\\
5	0.0801	\\
7.5	0.0507	\\
10	0.033	\\
12.5	0.0234	\\
15	0.0184	\\
17.5	0.0157	\\
20	0.0143	\\
22.5	0.0136	\\
25	0.0132	\\
27.5	0.013	\\
30	0.0129	\\
};

% MSM

%\addplot+[smooth,color=blue,dashed, every mark/.append style={solid, %fill=blue!50},mark=o,
%y filter/.code={\pgfmathparse{\pgfmathresult-0}\pgfmathresult}]
%  table[row sep=crcr]{%
%-10	0.3755	\\
%-7.5	0.3372	\\
%-5	0.2901	\\
%-2.5	0.2352	\\
%0	0.1766	\\
%2.5	0.1214	\\
%5	0.0769	\\
%7.5	0.0465	\\
%10	0.0285	\\
%12.5	0.0187	\\
%15	0.0135	\\
%17.5	0.0108	\\
%20	0.0093	\\
%22.5	0.0084	\\
%25	0.0078	\\
%27.5	0.0074	\\
%30	0.0072	\\
%};

%\addplot[smooth,color=green,solid,mark=none,
%y filter/.code={\pgfmathparse{\pgfmathresult-0}\pgfmathresult}]
%  table[row sep=crcr]{%
%	1 2\\
%};\label{P0}

\addplot[smooth,color=black,solid,mark=square,
y filter/.code={\pgfmathparse{\pgfmathresult-0}\pgfmathresult}]
  table[row sep=crcr]{%
	1 2\\
};\label{P1}

\addplot[smooth,color=black,solid, mark=diamond,
y filter/.code={\pgfmathparse{\pgfmathresult-0}\pgfmathresult}]
  table[row sep=crcr]{%
	1 2\\
};\label{P2}
%
%\addplot[smooth,color=cyan,solid,mark=none,
%y filter/.code={\pgfmathparse{\pgfmathresult-0}\pgfmathresult}]
%  table[row sep=crcr]{%
%	1 2\\
%};\label{P3}
%
%\addplot[smooth,color=blue,solid,mark=o,
%y filter/.code={\pgfmathparse{\pgfmathresult-0}\pgfmathresult}]
%  table[row sep=crcr]{%
%	1 2\\
%};\label{P4}
%
%\addplot[smooth,color=gray,solid,mark=none,
%y filter/.code={\pgfmathparse{\pgfmathresult-0}\pgfmathresult}]
%  table[row sep=crcr]{%
%	1 2\\
%};\label{P5}

\addplot[smooth,color=black,solid,mark=triangle,
y filter/.code={\pgfmathparse{\pgfmathresult-0}\pgfmathresult}]
  table[row sep=crcr]{%
	1 2\\
};\label{P8}

\node [draw,fill=white,font=\tiny,anchor= north  east] at (axis cs: 30,1) {
\setlength{\tabcolsep}{0.5mm}
\renewcommand{\arraystretch}{.8}
\begin{tabular}{l}
%\ref{P0}{Max Min total power constr.~\cite{Mo_2015}}\\
%\ref{P0}{PoPQ}\\
%\ref{P6}{Proposed PoP-8-PSK}\\
\ref{P8}{ZF-P phase quantized \cite{ZF-Precoding}}\\
%\ref{P4}{MSM Mapped \cite{MSM_precoder}}\\
\ref{P3}{CVX-CIO phase quantized \cite{CVX-CIO}}\\
\ref{P1}{Proposed MMDDT B\&B phase quantized}\\
\ref{P2}{MMDDT per antenna power constr.}\\
\end{tabular}
};

\end{axis}

\end{tikzpicture}%
%\vspace{-0.5em}
\captionsetup{justification=centering}
\caption{Uncoded BER versus $\mathrm{SNR}$, $K=2$, $\alpha_s=4$ and $\alpha_x=3$} 
\label{fig:BER_1}       % Give a unique label
%\vspace{-1em}
\end{center}
\end{figure}

The numerical computations were made with $K=2$ users, and the number of antennas at the BS is $M=6$ and $M=9$ and 1000 random channel realizations.  
One conventional configuration is considered with 8-PSK symbols ($\alpha_x=8$, $\alpha_s=8$). In addition, to demonstrate the flexibility of the proposed framework, a more exotic configuration is considered where $x_i$ is a 3-PSK symbol using QPSK modulation at the same time ($\alpha_x=3$, $\alpha_s=4$), which is compatible only with a subset of the existing methods.
The corresponding BER performances are illustrated in Fig.~\ref{fig:BER_1} and Fig.~\ref{fig:BER_2}, respectively.

The proposed method is compared with the following methods from the literature:
1. The MSM-Precoder \cite{MSM_precoder}, which corresponds to solving an LP with computational complexity in the order of $\mathcal{O}((2M+1)^{3.5})$, when using interior point methods (IPM);
2. The ZF precoder with constant envelope \cite{ZF-Precoding} with 
$\mathcal{O}(K^{2}M)$, which precoding vectors are subsequently phase quantized;
3. The CIO precoder implemented via CVX \cite{CVX-CIO}, which corresponds to solving a second order cone program with 
$\mathcal{O}((2M+1)^{3.5})$, when using IPM.
In addition, the Max-Min DDT precoder with full resolution and per antenna power constrained is considered, which yields a higher optimal value $\epsilon$ value, because relaxation of the feasible set results in an upper bound of the optimal value of the original problem. 
As expected the proposed algorithm shows a significantly lower BER than existing suboptimal algorithms, which confirms the aptitude of the Max Min DDT design objective  in the context of hard detection.

Note that the proposed algorithm does not yield an error floor which occurs for suboptimal precoding algorithms with phase quantization at the BS. 

%The MSM-Precoder is equivalent to just performing our algorithm initialization step, this produces then a sub optimal lower complexity solution. The continuous approach generates only a slightly better performance, only having significant gain of performance for SNR's greater than 25 dB.  

\begin{figure}[t]
%\vspace{-2em}
\begin{center}
%\sidecaption[t]
% This file was created by matlab2tikz v0.4.7 running on MATLAB 8.3.
% Copyright (c) 2008--2014, Nico Schlömer <nico.schloemer@gmail.com>
% All rights reserved.
% Minimal pgfplots version: 1.3
% 
% The latest updates can be retrieved from
%   http://www.mathworks.com/matlabcentral/fileexchange/22022-matlab2tikz
% where you can also make suggestions and rate matlab2tikz.
% 
%
% defining custom colors
\usetikzlibrary{positioning,calc}

\definecolor{mycolor1}{rgb}{0.00000,1.00000,1.00000}%
\definecolor{mycolor2}{rgb}{1.00000,0.00000,1.00000}%

\pgfplotsset{every axis label/.append style={font=\footnotesize},
every tick label/.append style={font=\footnotesize}
}

\begin{tikzpicture}[font=\footnotesize] 

\begin{axis}[%
name=A ,
ymode=log,
width  = 0.3\columnwidth,%5.63489583333333in,
%width  = 0.4\columnwidth,%5.63489583333333in,
%height = 0.3\columnwidth,%4.16838541666667in,
height = 0.4\columnwidth,%4.16838541666667in,
scale only axis,
xmin  = 3,
xmax  = 20,
xlabel= {$M$},
xmajorgrids,
ymin=5,
ymax=10^12,
ylabel={$\#_\textrm{branches}$ },
ymajorgrids,
%legend entries={$\alpha_x=3$,
%				$\alpha_x=4$,
				%1-bit B\&B    \cite{Landau2017}
                %%$M=14$,
                %%$M=16$,
%                },
%ytick={0, 0.5 , 1, 1.5, 2, 2.5 ,3,  3.1698  ,  4},
legend style={at={(0,1)},anchor=north west,draw=black,fill=white,legend cell align=left,font=\tiny}
]

\addlegendimage{color=black,fill=gray!20,mark=none}
\addlegendimage{color=red,dashed,fill=gray!20,mark=none}
%\addlegendimage{only marks,color=black,fill=green!20,mark=triangle}
%\addlegendimage{only marks,color=black,fill=red!20,mark=o}
%\addlegendimage{color=black,fill=gray!20,mark=diamond}

%FOR THE FIRST VALUE OF M
%PoPQ
%\addplot+[smooth,color=green,solid, every mark/.append style={solid, fill=gray!20},mark=square,
%y filter/.code={\pgfmathparse{\pgfmathresult-0}\pgfmathresult}]
%  table[row sep=crcr]{%
%   0  0.013200856250000\\
%   1  0.012471784615385\\
%   2  0.011701737019231\\
%   3  0.010904155769231\\
%   4  0.010084902884615\\
%   5  0.009252383173077\\
%   6  0.008428209615385\\
%   7  0.007620877884615\\
%   8  0.006853329807692\\
%   9  0.006140354807692\\
%   10 0.005483773557692\\
%   11 0.004902665384615\\
%   12 0.004395031730769\\
%   13 0.003964064423077\\
%   14 0.003599219711538\\
%   15 0.003298030769231\\
%   16 0.003057620673077\\
%   17 0.002862022115385\\
%   18 0.002710360576923\\
%   19 0.002588457211538\\
%   20 0.002494077884615\\
%   21 0.002421959134615\\
%   22 0.002364371634615\\
%   23 0.002320342788462\\
%   24 0.002284522115385\\
%   25 0.002255893750000\\
%};

%Exhaustive Search alpha_x=3
\addplot+[smooth,color=black,solid, every mark/.append style={solid, fill=cyan!20},mark=asterisk,
y filter/.code={\pgfmathparse{\pgfmathresult-0}\pgfmathresult}]
  table[row sep=crcr]{%
3	27	\\
4	81	\\
5	243	\\
6	729	\\
7	2187	\\
8	6561	\\
9	19683	\\
10	59049	\\
11	177147	\\
12	531441	\\
13	1594323	\\
14	4782969	\\
15	14348907	\\
16	43046721	\\
17  129100000   \\
18  387400000   \\
19  1162300000  \\
20  3486800000  \\
};

% max min bb - alpha x 3
\addplot+[smooth,color=black,solid, every mark/.append style={solid, fill=blue!20},mark=triangle,
y filter/.code={\pgfmathparse{\pgfmathresult-0}\pgfmathresult}]
  table[row sep=crcr]{%
3	7.365	\\
4	13.4932	\\
5	21.3622	\\
6	31.002	\\
7	44.1397	\\
8	59.3835	\\
9	77.7982	\\
10	99.5482	\\
11	123.8047	\\
12	155.5582	\\
13	189.6428	\\
14	230.664	\\
15	269.178	\\
16	325.026	\\
17	395.397	\\
18	454.3785	\\ 
19	507.1	\\
20	589.9605	\\
};

%1-bit Landau 2017
\addplot+[smooth,color=red,dashed, every mark/.append style={solid, fill=cyan!20},mark=diamond,
y filter/.code={\pgfmathparse{\pgfmathresult-0}\pgfmathresult}]
  table[row sep=crcr]{%
%\\\\\2	7.4625	\\
3	14.678	\\
4	24.905	\\
5	38.1737	\\
6	54.4928	\\
7	76.2362	\\
8	105.7435	\\
9	137.3947	\\
10	174.643	\\
11	224.3965	\\
12	282.6578	\\
13	347.003	\\
14	422.882	\\
15 510.5582 \\
16	617.3265	\\
17	748.751	\\
18	873.0092	\\
19	1018.5	\\
20	1154.7	\\
};

%Exhaustive Search alpha_x=4
\addplot+[smooth,color=red, dashed, every mark/.append style={solid, fill=cyan!20},mark=asterisk,
y filter/.code={\pgfmathparse{\pgfmathresult-0}\pgfmathresult}]
  table[row sep=crcr]{%
2	16	\\
3	64	\\
4	256	\\
5	1024	\\
6	4096	\\
7	16384	\\
8	65536	\\
9	262144	\\
10	1048576	\\
11	4194304	\\
12	16777216	\\
13	67108864	\\
14	268435456	\\
15	1073741824	\\
16	4294967296	\\
17  17200000000 \\
18  68700000000 \\
19  274900000000 \\
20  1099500000000 \\
};

%\addplot[smooth,color=red,dashed,
%y filter/.code={\pgfmathparse{\pgfmathresult-0}\pgfmathresult}]
%  table[row sep=crcr]{%
%	-20 2\\
%};\label{P99}
%
%\addplot[smooth,color=black,solid,
%y filter/.code={\pgfmathparse{\pgfmathresult-0}\pgfmathresult}]
%  table[row sep=crcr]{%
%	-20 2\\
%};\label{P88}
%
%
%
%
%
%
%
%\node [draw,fill=white,font=\tiny,anchor= north  east] at (axis cs: %20,0.7) {
%\setlength{\tabcolsep}{0.5mm}
%\renewcommand{\arraystretch}{.8}
%\begin{tabular}{l}
%\ref{P99}{$\alpha_x=4$}\\
%\ref{P88}{$\alpha_x=3$}\\
%\end{tabular}
%};

\end{axis}

\begin{axis}[%
name=SumRate,
at={($(A.east)+(40,0em)$)},
                anchor= west,
name=B ,
ymode=log,
width  = 0.3\columnwidth,%5.63489583333333in,
%width  = 0.4\columnwidth,%5.63489583333333in,
%height = 0.3\columnwidth,%4.16838541666667in,
height = 0.4\columnwidth,%4.16838541666667in,
scale only axis,
xmin  = 3,
xmax  = 20,
xlabel= {$M$},
xmajorgrids,
ymin=7,
ymax=10000,
ylabel={$\#_\textrm{branches}$ },
ymajorgrids,
legend entries={Ex. Search,
				B\&B - $\alpha_x=3$,
				1-bit B\&B    \cite{Landau2017}
                %$M=14$,
                %$M=16$,
                },
legend style={at={(0,1)},anchor=north west,draw=black,fill=white,legend cell align=left,font=\tiny}
]

\addlegendimage{color=black,fill=gray!20,mark=asterisk}
\addlegendimage{color=black,fill=gray!20,mark=triangle}
%\addlegendimage{only marks,color=black,fill=green!20,mark=triangle}
%\addlegendimage{only marks,color=black,fill=red!20,mark=o}
\addlegendimage{color=black,fill=gray!20,mark=diamond}

%\addlegendimage{color=black,fill=gray!20,mark=asterisk}
%\addlegendimage{color=black,fill=gray!20,mark=triangle}
%\addlegendimage{only marks,color=black,fill=green!20,mark=triangle}
%\addlegendimage{only marks,color=black,fill=red!20,mark=o}
%\addlegendimage{color=black,fill=gray!20,mark=square}

%FOR THE FIRST VALUE OF M
%PoPQ
%\addplot+[smooth,color=green,solid, every mark/.append style={solid, fill=gray!20},mark=square,
%y filter/.code={\pgfmathparse{\pgfmathresult-0}\pgfmathresult}]
%  table[row sep=crcr]{%
%   0  0.013200856250000\\
%   1  0.012471784615385\\
%   2  0.011701737019231\\
%   3  0.010904155769231\\
%   4  0.010084902884615\\
%   5  0.009252383173077\\
%   6  0.008428209615385\\
%   7  0.007620877884615\\
%   8  0.006853329807692\\
%   9  0.006140354807692\\
%   10 0.005483773557692\\
%   11 0.004902665384615\\
%   12 0.004395031730769\\
%   13 0.003964064423077\\
%   14 0.003599219711538\\
%   15 0.003298030769231\\
%   16 0.003057620673077\\
%   17 0.002862022115385\\
%   18 0.002710360576923\\
%   19 0.002588457211538\\
%   20 0.002494077884615\\
%   21 0.002421959134615\\
%   22 0.002364371634615\\
%   23 0.002320342788462\\
%   24 0.002284522115385\\
%   25 0.002255893750000\\
%};

% max min bb - alpha x 3
\addplot+[smooth,color=black,solid, every mark/.append style={solid, fill=blue!20},mark=triangle,
y filter/.code={\pgfmathparse{\pgfmathresult-0}\pgfmathresult}]
  table[row sep=crcr]{%
3	7.365	\\
4	13.4932	\\
5	21.3622	\\
6	31.002	\\
7	44.1397	\\
8	59.3835	\\
9	77.7982	\\
10	99.5482	\\
11	123.8047	\\
12	155.5582	\\
13	189.6428	\\
14	230.664	\\
15	269.178	\\
16	325.026	\\
17	395.397	\\
18	454.3785	\\ 
19	507.1	\\
20	589.9605	\\
};

%1-bit Landau 2017
\addplot+[smooth,color=red,dashed, every mark/.append style={solid, fill=cyan!20},mark=diamond,
y filter/.code={\pgfmathparse{\pgfmathresult-0}\pgfmathresult}]
  table[row sep=crcr]{%
%\\\\\2	7.4625	\\
3	14.678	\\
4	24.905	\\
5	38.1737	\\
6	54.4928	\\
7	76.2362	\\
8	105.7435	\\
9	137.3947	\\
10	174.643	\\
11	224.3965	\\
12	282.6578	\\
13	347.003	\\
14	422.882	\\
15 510.5582 \\
16	617.3265	\\
17	748.751	\\
18	873.0092	\\
19	1018.5	\\
20	1154.7	\\
};

\addplot[smooth,color=red,dashed,
y filter/.code={\pgfmathparse{\pgfmathresult-0}\pgfmathresult}]
  table[row sep=crcr]{%
	-20 2\\
};\label{P99}
\addplot[smooth,color=black,solid,
y filter/.code={\pgfmathparse{\pgfmathresult-0}\pgfmathresult}]
  table[row sep=crcr]{%
	-20 2\\
};\label{P88}

\node [draw,fill=white,font=\tiny,anchor= south  east] at (axis cs: 20,7) {
\begin{tabular}{l}
\ref{P99}{$\alpha_x=4$}\\
\ref{P88}{$\alpha_x=3$}\\
\end{tabular}
};

\end{axis}

\end{tikzpicture}%
%\vspace{-0.5em}
\captionsetup{justification=centering}
\caption{Average $\#$ of branches visited versus Number of transmit antennas, $K=2$} 
\label{fig:Complexity}       % Give a unique label
%\vspace{-1em}
\end{center}
\end{figure}
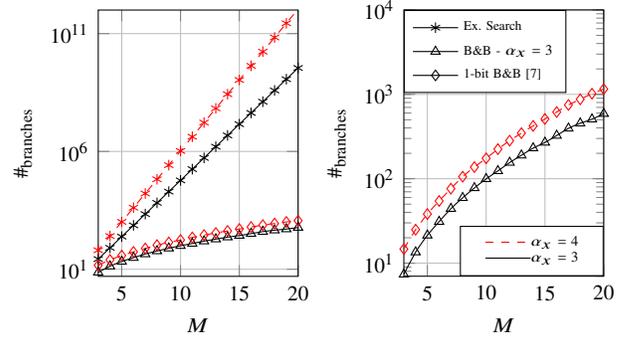

 The proposed branch-and-bound method yields the same solution as the exhaustive search but with a lower average complexity.
 The complexity of the algorithm heavily depends on finding as early as possible a tight upper bound that permits many exclusions of possible candidates while going down the tree. 
%In the most pessimistic case the problem is NP-hard, however the average %complexity is polynomial.
By using IPM for solving sub problems \eqref{eq:bb_optimzation_problem} corresponds to a computational complexity given by $\mathcal{O}(n^{3.5})$, with $n\leq(2M+1)$.
Note that the dimensions of the sub problems decrease when climbing down the tree.
The average number of sub problems is illustrated in Fig.~\ref{fig:Complexity}. 
Based on Fig.~\ref{fig:Complexity} it turns out that the average number of sub problems is only a small fraction of the number of candidates which are evaluated in the exhaustive search. Taking into account that each candidate evaluation in the exhaustive search corresponds to a complexity of $\mathcal{O}(MK)$  
justifies the utilization of the proposed branch-and-bound approach, when the optimal precoding vector is desired. 

\section{Conclusions}

An optimal algorithm for precoding constrained to constant envelope and phase quantization for PSK modulation and hard detection is proposed. The design criterion maximizes the minimum distance to the decision threshold at the receivers. The proposed algorithm
outperforms the state-of-art techniques for this class of precoding in terms of BER. Numerical results confirm the efficiency of the proposed branch-and-bound strategy.

\bibliographystyle{IEEEtran}
\bibliography{bib-refs}

\begin{thebibliography}{10}
\providecommand{\url}[1]{#1}
\csname url@samestyle\endcsname
\providecommand{\newblock}{\relax}
\providecommand{\bibinfo}[2]{#2}
\providecommand{\BIBentrySTDinterwordspacing}{\spaceskip=0pt\relax}
\providecommand{\BIBentryALTinterwordstretchfactor}{4}
\providecommand{\BIBentryALTinterwordspacing}{\spaceskip=\fontdimen2\font plus
\BIBentryALTinterwordstretchfactor\fontdimen3\font minus
  \fontdimen4\font\relax}
\providecommand{\BIBforeignlanguage}[2]{{%
\expandafter\ifx\csname l@#1\endcsname\relax
\typeout{** WARNING: IEEEtran.bst: No hyphenation pattern has been}%
\typeout{** loaded for the language `#1'. Using the pattern for}%
\typeout{** the default language instead.}%
\else
\language=\csname l@#1\endcsname
\fi
#2}}
\providecommand{\BIBdecl}{\relax}
\BIBdecl

\bibitem{Walden_1999}
R.~Walden, ``Analog-to-digital converter survey and analysis,'' \emph{IEEE J.
  Sel. Areas Commun.}, vol.~17, no.~4, pp. 539 --550, Apr. 1999.

\bibitem{ZF-Precoding}
S.~K. {Mohammed} and E.~G. {Larsson}, ``Per-antenna constant envelope precoding
  for large multi-user {MIMO} systems,'' \emph{{{IEEE} Trans. Commun.}},
  vol.~61, no.~3, pp. 1059--1071, March 2013.

\bibitem{MMSE}
A.~{Mezghani}, R.~{Ghiat}, and J.~A. {Nossek}, ``Transmit processing with low
  resolution {D/A}-converters,'' in \emph{Proc.\ of the 16th IEEE Int. Conf. on
  Electronics, Circuits and Systems - (ICECS 2009)}, Hammamet,Tunisia, Dec
  2009, pp. 683--686.

\bibitem{Jacobsson2018}
S.~{Jacobsson}, W.~{Xu}, G.~{Durisi}, and C.~{Studer}, ``{MSE}-optimal 1-bit
  precoding for multiuser {MIMO} via branch and bound,'' in \emph{Proc.\ of
  2018 IEEE International Conference on Acoustics, Speech and Signal Processing
  (ICASSP)}, Calgary, Alberta, Canada, April 2018, pp. 3589--3593.

\bibitem{Landau_SCC2013}
L.~Landau, S.~Krone, and G.~P. Fettweis, ``Intersymbol-interference design for
  maximum information rates with 1-bit quantization and oversampling at the
  receiver,'' in \emph{Proc. of the Int. ITG Conf. on Systems, Communications
  and Coding}, Munich, Germany, Jan. 2013.

\bibitem{Mo_2015}
J.~Mo and R.~W. Heath~Jr, ``Capacity analysis of one-bit quantized {MIMO}
  systems with transmitter channel state information,'' \emph{{{IEEE} Trans.
  Signal Process.}}, vol.~63, no.~20, pp. 5498--5512, Oct 2015.

\bibitem{Landau2017}
L.~T.~N. {Landau} and R.~C. {de Lamare}, ``Branch-and-bound precoding for
  multiuser {MIMO} systems with 1-bit quantization,'' \emph{{{IEEE} Wireless
  Commun. Lett.}}, vol.~6, no.~6, pp. 770--773, Dec 2017.

\bibitem{MSM_precoder}
H.~{Jedda}, A.~{Mezghani}, A.~L. {Swindlehurst}, and J.~A. {Nossek},
  ``Quantized constant envelope precoding with {PSK} and {QAM} signaling,''
  \emph{{{{IEEE} Trans. Wireless Commun.}}}, vol.~17, no.~12, pp. 8022--8034,
  Dec 2018.

\bibitem{Boyd_2004}
S.~Boyd and L.~Vandenberghe, \emph{Convex Optimization}.\hskip 1em plus 0.5em
  minus 0.4em\relax New York, NY, USA: Cambridge University Press, 2004.

\bibitem{Jedda_2016}
H.~Jedda, J.~A. Nossek, and A.~Mezghani, ``Minimum {BER} precoding in 1-bit
  massive {MIMO} systems,'' in \emph{Proc. of IEEE Sensor Array and
  Multichannel Signal Processing Workshop (SAM)}, Rio de Janeiro, Brazil, July
  2016.

\bibitem{CVX-CIO}
P.~V. {Amadori} and C.~{Masouros}, ``Constant envelope precoding by
  interference exploitation in phase shift keying-modulated multiuser
  transmission,'' \emph{{{IEEE} Trans. Commun.}}, vol.~16, no.~1, pp. 538--550,
  Jan 2017.

\end{thebibliography}

\end{document}